\documentclass[usegraphicx,useAMS,usenatbib]{mn2e} 
\usepackage{amssymb}
\usepackage{times}
\usepackage{aas_macros} 

\usepackage[usenames]{color}

\newcommand{\ione}{\,{\sc i}}
\newcommand{\itwo}{\,{\sc ii}}

\newcommand{\vwa}{{VWA}}

\newcommand{\vmicro}{{microturbulence}}

\newcommand{\teff}{$T_{\rm eff}$}

\newcommand{\logg}{$\log g$}

\newcommand{\feh}{[Fe/H]}

\newcommand{\vsini}{$v \sin i$}
\newcommand{\kms}{km\,s$^{-1}$}

\newcommand{\feone}{Fe\,{\sc i}}
\newcommand{\fetwo}{Fe\,{\sc ii}}

\newcommand{\corot}{{\em CoRoT}}

\newcommand{\kepler}{{\em Kepler}}

\title[Accurate parameters of 93 solar-type {\it Kepler} targets]
{Accurate fundamental parameters and detailed abundance patterns from
spectroscopy of 93 solar-type {\it Kepler} targets\thanks{Based on observations obtained at the Canada-France-Hawaii
Telescope (CFHT) which is operated by the National Research
Council of Canada, the Institut National des Sciences de l'Univers
of the Centre National de la Recherche Scientifique of France, and
the University of Hawaii.}\thanks{Based on observations with the 2-m Telescope Bernard Lyot funded by the 
CNRS Institut National des Sciences de l'Univers.}}

\author[H. Bruntt et al.]
{
  H.~Bruntt$^{1,2}$\thanks{E-mail: bruntt@gmail.com},
  S.~Basu$^{3}$,
  B.~Smalley$^{4}$,
  W.~J.~Chaplin$^{5}$,
  G.~A.~Verner$^{5,6}$,
  T.~R.~Bedding$^{7}$,\newauthor
  C.~Catala$^{2}$,
  J.-C.~Gazzano$^{8}$,
  J.~Molenda-\.Zakowicz$^{9}$,
  A.~O.~Thygesen$^{1}$
  K.~Uytterhoeven$^{10,17}$,\newauthor
  S.~Hekker$^{11,5}$,
  D.~Huber$^{7}$,
  C.~Karoff$^{1}$,
  S.~Mathur$^{12}$,
  B.~Mosser$^{2}$,
  T.~Appourchaux$^{13}$,\newauthor
  T.~L.~Campante$^{14,1}$,
  Y.~Elsworth$^{5}$,
  R.~A.~Garc\'ia$^{15}$,
  R.~Handberg$^{1}$,
  T.~S.~Metcalfe$^{12}$,\newauthor
  P.-O.~Quirion$^{16}$,
  C.~R\'egulo$^{10,17}$,
  I.~W.~Roxburgh$^{6}$,
  D.~Stello$^{7}$,
  J.~Christensen-Dalsgaard$^{1}$,\newauthor
  S.~D.~Kawaler$^{18}$,
  H.~Kjeldsen$^{1}$,
  R.~L.~Morris$^{19}$,
  E.~V.~Quintana$^{19}$,
  D.~T.~Sanderfer$^{20}$\\
\\
$^{1}$Department of Physics and Astronomy, Aarhus University, DK-8000
Aarhus C, Denmark\\
$^{2}$LESIA, CNRS, Universit\'e Pierre et Marie Curie, Universit\'e
Denis Diderot, Observatoire de Paris, 92195 Meudon Cedex, France\\
$^{3}$Department of Astronomy, Yale University, P.O. Box 208101, New
Haven, CT 06520-8101, USA\\
$^{4}$Astrophysics Group, Keele University, Staffordshire, UK, ST5 5BG\\
$^{5}$School of Physics and Astronomy, University of Birmingham,
Edgbaston, Birmingham, B15 2TT, UK\\
$^{6}$Astronomy Unit, Queen Mary, University of London, Mile End Road,
London, E1 4NS, UK\\
$^{7}$Sydney Institute for Astronomy (SIfA), School of Physics,
University of Sydney, NSW 2006, Australia\\
$^{8}$Laboratoire d'Astrophysique de Marseille (UMR 6110), OAMP,
Universit\'e Aix-Marseille \& CNRS, 38 rue Fr\'ed\'eric Joliot Curie,
13388 Marseille cedex 13, France\\
$^{9}$Astronomical Institute, University of Wroc\l{}aw, ul. Kopernika,
11, 51-622 Wroc\l{}aw, Poland\\
$^{10}$Instituto de Astrof\'{\i}sica de Canarias, E-38200 La Laguna,
Tenerife, Spain\\
$^{11}$Astronomical Institute, `Anton Pannekoek', University of
Amsterdam, PO Box 94249, 1090 GE Amsterdam, The Netherlands\\
$^{12}$High Altitude Observatory,
National Center for Atmospheric Research, Boulder, Colorado 80307, USA\\
$^{13}$Institut d'Astrophysique Spatiale, Universit\'e Paris XI -- CNRS
(UMR8617), Batiment 121, 91405 Orsay Cedex, France\\
$^{14}$Centro de Astrof\'isica, Universidade do Porto, Rua das
Estrelas, 4150-762 Porto, Portugal\\
$^{15}$Laboratoire AIM, CEA/DSM -- CNRS -- Universit\'e Paris Diderot
-- IRFU/SAp, 91191 Gif-sur-Yvette Cedex, France\\
$^{16}$Canadian Space Agency, 6767 Boulevard de l'A\'eroport,
Saint-Hubert, QC, J3Y 8Y9, Canada.\\
$^{17}$Departamento de Astrof\'{\i}sica, Universidad de La Laguna,
E-38206 La Laguna, Tenerife, Spain\\
$^{18}$Department of Physics and Astronomy, Iowa State University,
Ames, IA 50011, USA\\
$^{19}$SETI Institute/NASA Ames Research Center, Moffett Field, CA 94035,
USA\\
$^{20}$NASA Ames Research Center, Moffett Field, CA 94035, USA
}

\begin{document}

\date{Accepted 2011 MONTH DATE. Received 2011 MONTH DATE; in original form 2011 MONTH DATE}

\pagerange{\pageref{firstpage}--\pageref{lastpage}} \pubyear{2011}

\maketitle

\label{firstpage}

\begin{abstract}

We present a detailed spectroscopic study of 93 solar-type stars  that are
targets of the NASA/\kepler\ mission and  provide detailed chemical composition
of each target.  We find that the overall metallicity is well-represented by Fe
lines. Relative abundances of light elements (CNO) and $\alpha$~elements are
generally higher for  low-metallicity stars. Our spectroscopic analysis {
benefits} from the accurately measured surface gravity from the asteroseismic
analysis of the \kepler\ light curves.  The $\log g$ parameter is known to
better than 0.03 dex and is held fixed in the analysis.  We compare our \teff\
determination with a recent colour calibration of $V_{\rm T}-K_{\rm S}$ (TYCHO
$V$ magnitude minus 2MASS $K_{\rm S}$ magnitude) and find very good agreement
and a scatter of only 80~K, showing that for other nearby \kepler\ targets  this
index can be used. The asteroseismic \logg\ values agree very well with the
classical determination using \feone-\fetwo\ balance,  although we find a small
systematic offset of $0.08$ dex (asteroseismic \logg\ values are lower).   The
abundance patterns of metals, $\alpha$ elements, and the light elements (CNO)
show that a simple scaling by [Fe/H] is adequate to represent the metallicity of
the stars, except for the  stars  with metallicity below $-0.3$, where
$\alpha$-enhancement becomes important. However, this  is only important for a
very small fraction of the \kepler\ sample. We therefore recommend that a simple
scaling with [Fe/H] be employed in the asteroseismic analyses of large ensembles
of solar-type stars.




\end{abstract}


\section{Introduction}

The \kepler\ Space Mission, launched by NASA on 7 March 2009, has the goal of
detecting transiting exoplanets down to Earth-size around solar-type stars
\citep{2010Sci...327..977B}. The photometric transit depth provides a direct
measure of the ratio of the planet-to-star radii. Hence, accurate measurements
of the absolute planet radii depend on having robust measurements of the stellar
radii. While ``classical'' methods of stellar radius determination suffer from
large uncertainties, radii can be determined to a few percent from asteroseismic
modelling of the averaged stellar pulsation data
\citep{2009ApJ...700.1589S,gai11}. Thus, the exquisite \kepler\ data
\citep{2010ApJ...713L.160G,2010ApJ...713L.120J} fuel an interesting
synergy between exoplanet science and asteroseismology.

Asteroseismology is a powerful tool to sound the stellar interiors and determine
their basic properties, such as radius, mass and age (see e.g.,
\citealt{2010aste.book.....A,2011arXiv1104.5191C} and references therein). Much
recent progress has been made with ground-based spectroscopy and the \corot\
photometry space missions, with solar-like oscillations detected in a few dozen
stars \citep{michel08,2011arXiv1107.1723B}. The \kepler\ mission is a major step
forward, with oscillations having been detected during an asteroseismic survey
in several hundred solar-type stars \citep{chaplin11}. More than 150 of these
stars have since been observed continuously, the expectation
being that a sizable fraction will be monitored for the entire duration of
the mission. This will  make it possible to investigate stars with convective
cores and stars with extensive outer convective envelopes. Reliable measurements
of the stars' fundamental properties firmly constrain the range of models and
allows us to exploit the seismic data much more fully
\citep{1994ApJ...427.1013B,basu10}.  For example, the uncertainty in the
measurement of the stellar radius drops from about 40 percent to less than 3
percent when the average large frequency spacing (the frequency between mode
frequencies with the same degree and consecutive order of the pulsation modes)
is combined with spectroscopic measurements of \teff\ and metallicity ([Fe/H]).
The technique described by \cite{basu10} requires an accuracy in the
spectroscopic parameters better than about 200\,K in \teff\ for main sequence
stars and 0.2 dex in [Fe/H] for more evolved stars.  For the detailed theoretical
modelling, such accurate parameters are critical for removing the degeneracy
between mass and metallicity.

The \kepler\ Input Catalogue (KIC) \citep{2011AJ....142..112B} contains estimates of the
parameters of the stars, determined by calibrating $griz$ + Mg\,$b$ Sloan filter
photometry of all stars in the 105 square degree field observed by \kepler.  The
precision of the KIC parameters has been called into question by recent
spectroscopic analyses \citep{jmz11, bruntt10-gig}. It has been found that
\teff\ only agrees within about 300\,K, while  \logg\ can be off by 0.5 dex, and
[Fe/H] by up to 2.0 dex. The previous study by \cite{jmz11} reported
spectroscopic results on a wide range of stars, from solar-type to A-stars,
which relied on medium-resolution spectra and with relatively low
signal-to-noise ratio (S/N). In the current study we focus on the FGK 
(solar-type) stars, using spectra with high S/N (200 or more) and high
resolution ($R=80,000$). This enables us to measure accurate values for \teff\
and to obtain a detailed abundance pattern for each star. 

\section{Observations and data reduction}

The spectra were obtained with the ESPaDOnS spectrograph at  the 3.6-m
Canada-France-Hawaii Telescope (CFHT) \citep{donati06} in USA and  with the
NARVAL spectrograph mounted on  the 2-m Bernard Lyot Telescope at the Pic du
Midi Observatory in France. In both facilities the observations were carried out
as service observations from May to September in 2010.  The two spectropolarimetric
instruments are very similar and we used the ``star mode'' with a 
nominal resolution of $R = 80\,000$. 
The integration times were typically a few minutes up to 15 minutes 
and were adjusted to obtain S/N ratios in the range
200 to 300 per pixel.  Here, we report results for the 93 stars that have detections of
solar-type oscillations in the \kepler\ light curve data.  An additional 20
stars, which are evolved stars ($\log g < 3.5$), will be presented by 
Thygesen et~al.\ (in preparation). However,  these stars are included here in our calibration
of \teff\ and microturbulence.

We have used the standard pipeline-reduced spectra \citep{1997MNRAS.291..658D},
in which the overlapping part of the \'{e}chelle orders have not been merged. We
used the program called {\tt rainbow} \citep{bruntt10-corot} to normalize the
spectra manually. We carefully adjusted the continua by comparing with a
synthetic spectrum calculated with the program that was used for the spectral
line fitting. We made sure the overlapping part of the orders agreed on both the
line depths and the level of the continuum. A few of the stars were observed on
different nights with both spectrographs and we confirmed that their spectra
were very similar. In these cases we used the spectrum with the highest S/N
ratio in the continuum around 6000\,\AA. One star, KIC 8379927 (HD 187160), was
not  included in the analysis because it is an double-lined spectroscopic binary
\citep{2007Obs...127..313G}. In addition, two stars (KIC 12155015 and 8831759)
were found to be too cool ({\teff} $<$4500~K) for detailed analysis.

\section{Data analysis}

We used the semi-automatic  software \vwa\footnote{VWA is available at
https://sites.google.com/site/vikingpowersoftware/}  \citep{bruntt23} to derive
fundamental stellar parameters and elemental abundances from the spectra. This
software has been used to analyse several of the \corot\ exoplanet host stars
\citep{bruntt10-corot} and asteroseismic targets \citep{bruntt09-corot}, and the
results compared well with other methods, including  interferometric
determinations of the effective temperature \citep{bruntt23}. It has also been
demonstrated that the $\log g$ determination is accurate for nearby binary
systems \citep{bruntt23}, including Procyon and $\alpha$~Cen~A+B.

VWA treats each line individually and fits them by iteratively changing the
abundance and calculating synthetic profiles. VWA automatically takes into
account any line blending. This makes it possible to analyse the F-type stars in
the sample, which have wide lines with \vsini\ in the range 10 to 40\,\kms, as
well as  cool stars (\teff $>$ 4500~K) where blending becomes severe in the blue
part of the spectrum. We have extracted the atomic line lists from the VALD
database \citep{kupka99}, using the appropriate \teff\ and \logg\ of each star.
This is important in order for the line depths computed by VALD to be
approximately correct. In addition, we updated the atomic line parameters in the
region around the lithium line at 6707\,\AA,  using the line list from
\cite{ghezzi09}, but we note that our code cannot  take into account the weak CN
molecular bands.

VWA uses atmospheric models that are interpolated in a grid of  MARCS
models\footnote{Available at {http://marcs.astro.uu.se/}} \citep{gustafsson08} 
for a fixed microturbulence of 2\,\kms. We tested that changing to 1\,\kms\ had 
a negligible effect on the results. The grid had a mesh size of 250\,K for
\teff, 0.5 for \logg, and variable for [Fe/H], with a step size of typically
0.25~dex around the solar value. Abundances in MARCS models were all scaled
relative to the solar abundance from \cite{grevesse07}.

\begin{table}
 \centering
  \caption{Atomic data used in the spectra analysis.
The wavelength, excitation energy and 
oscillator strengths ($\log gf$) from VALD and the
adjusted values are given.
We list C,N, and O and the complete table is 
available in the electronic version.
\label{tab:loggf}}
  \begin{tabular}{rc rrr}
\hline \hline
    &                      & Excitation     & VALD      & Adjusted  \\
El. & $\lambda$\,$[$\AA$]$ &  $[$eV$]$      & $\log gf$ & $\log gf$ \\ \hline
C\,{\sc i}&$  4932.049  $&$  7.685  $&$ -1.884  $&$ -1.070  $\\
     &$  5052.167  $&$  7.685  $&$ -1.648  $&$ -1.196  $\\
     &$  5380.337  $&$  7.685  $&$ -1.842  $&$ -1.475  $\\
     &$  5800.602  $&$  7.946  $&$ -2.338  $&$ -2.075  $\\
     &$  6010.675  $&$  8.640  $&$ -4.605  $&$  0.031  $\\
     &$  6014.834  $&$  8.643  $&$ -1.585  $&$ -1.379  $\\
     &$  6397.961  $&$  8.771  $&$ -1.778  $&$ -1.287  $\\
     &$  6587.610  $&$  8.537  $&$ -1.596  $&$ -0.901  $\\
     &$  6655.517  $&$  8.537  $&$ -1.370  $&$ -1.669  $\\
     &$  7111.469  $&$  8.640  $&$ -2.140  $&$ -0.209  $\\
     &$  7113.179  $&$  8.647  $&$ -0.774  $&$ -0.700  $\\
     &$  7115.168  $&$  8.643  $&$ -0.935  $&$ -0.808  $\\
     &$  7116.988  $&$  8.647  $&$ -0.508  $&$  0.093  $\\
     &$  7119.657  $&$  8.643  $&$ -1.149  $&$ -0.785  $\\
     &$  8058.624  $&$  8.851  $&$ -1.275  $&$ -0.907  $\\
     &$  8335.148  $&$  7.685  $&$ -0.420  $&$ -0.117  $\\ \hline
 N\,{\sc i}&$  7442.300  $&$ 10.330  $&$ -0.384  $&$ -0.030  $\\
     &$  7468.313  $&$ 10.336  $&$ -0.189  $&$  0.137  $\\
     &$  8629.240  $&$ 10.690  $&$  0.075  $&$  0.670  $\\
     &$  8683.405  $&$ 10.330  $&$  0.086  $&$  0.527  $\\
     &$  8711.707  $&$ 10.330  $&$ -0.234  $&$  0.139  $\\ \hline
 O\,{\sc i}&$  5577.339  $&$  1.967  $&$ -8.204  $&$ -7.726  $\\
     &$  6156.776  $&$ 10.741  $&$ -0.694  $&$ -0.109  $\\
     &$  6158.186  $&$ 10.741  $&$ -0.409  $&$ -0.113  $\\
     &$  6300.304  $&$  0.000  $&$ -9.819  $&$-10.268  $\\
     &$  7002.229  $&$ 10.989  $&$ -1.855  $&$ -0.165  $\\
     &$  7771.941  $&$  9.146  $&$  0.369  $&$  0.820  $\\
     &$  7774.161  $&$  9.146  $&$  0.223  $&$  0.642  $\\
     &$  7775.390  $&$  9.146  $&$  0.001  $&$  0.357  $\\
\hline
\end{tabular}
\end{table}

\subsection{Analysis of solar spectra}


VWA relies on a differential approach, with a solar spectrum as the reference,
in order to correct for systematics in the atmospheric models and to minimize
errors in the atomic oscillator strengths. We adopted the new version of the
Fourier Transform Spectrometer (FTS) Kitt Peak Solar Flux Atlas
\citep{kurucz06}, which was also adopted by \cite{asplund09}.  This spectrum is
a new reduction of the original spectrum obtained by \cite{kurucz84}.   When we
compared the FTS spectrum with the calculated spectrum for the  canonical solar
values of $T_{\rm eff} = 5777$\,K and $\log g = 4.437$,  we found that the
continuum of the synthetic profiles lies at 1.0 but the FTS spectrum is
generally somewhat lower, especially in the blue part of the spectrum.  This is
probably due to the many ``missing lines'' { in present-day line
compilations.} Since abundances are derived from the synthetic spectra, we
re-normalized the FTS spectrum to have the same continuum level (close to 1.0)
as the synthetic spectrum of the Sun. We used the same normalization approach
for all the \kepler\ targets.


To remove instrumental effects, one option would be to use
a solar spectrum  observed with the same instrument as for the observations. The
reason for adopting the FTS spectrum is its superior quality compared to
what can be obtained through a normal calibration spectrum with an \'{e}chelle
spectrograph. The FTS spectrum  has a resolution of about $R=300\,000$ and a S/N
of about 3000.  We were thus able to determine oscillator strength ($\log gf$) corrections 
for more than 1200 lines  in the range 4500 to 9200\,\AA. Secondly, the FTS instrument
design serves to minimize scattered light,  which is known to affect \'{e}chelle
spectra. In Table~\ref{tab:loggf} we list atomic data for lines
that were used in at least 10 of the \kepler\ targets. The complete
table is available in the electronic version.

\begin{table}
 \centering
  \caption{Properties of observed solar spectra.
The $\Delta {\rm [Fe/H]}$ is the mean offset in Fe abundance
between the FTS spectrum and the other spectra and $N$
is the number of spectral lines used.
\label{tab:sun}}
  \begin{tabular}{llcc}
\hline
   Spectrograph & Source of light  & $\Delta {\rm [Fe/H]}$ & $N$ \\ \hline
FTS             & Direct           & $\equiv 0.00$    & 405 \\
ESPaDOnS        & Twilight Sky     & $ -0.04\pm0.04$  & 284 \\
NARVAL          & Moon             & $ -0.03\pm0.04$  & 276 \\
HARPS           & Moon             & $ -0.02\pm0.04$  & 197 \\
HARPS           & Ceres            & $ -0.02\pm0.03$  & 214 \\
FIES            & Day Sky          & $ -0.05\pm0.06$  & 280 \\
HERMES          & Venus            & $ -0.06\pm0.07$  & 347 \\
\hline
\end{tabular}
\end{table}

To investigate the influence of the adopted FTS solar spectrum, we analysed
several  spectra of the Sun taken with the spectrographs used for the \kepler\
stars: NARVAL and ESPaDOnS. In addition, we also analysed solar spectra from
different spectrographs, including  HARPS at the La Silla 3.6-m telescope, FIES
at the 2.56-m Nordic Optical Telescope, and HERMES at the 1.2-m MERCATOR
telescope. The solar spectra we acquired by observing scattered sunlight either
from twilight or daytime sky, or by observing the Moon, Ceres or Venus. 

In Table~\ref{tab:sun} we list the seven solar spectra we have analysed. 
We used lines in the range 4500 to 6800\,\AA\ and equivalent widths from 5 to 105\,m\AA. 
Typically between 200 and 300 lines were used in each spectrum, depending on the S/N. 
When measuring abundances of each line relative to the FTS spectrum, we found a very
slight correlation with wavelength, presumably due to insufficient removal of
scattered light. In addition, we found that the spectral lines were weaker in
all spectra compared to the FTS, giving lower abundances by 0.03 -- 0.07 dex. 
In Table~\ref{tab:sun} we list the offset for \feone\ and note that offsets
are nearly identical for the other elements. This is expected if the cause is
dilution of the lines by scattered light. We recommend that in future analyses, 
a spectrum of the Sun is always acquired to estimate the maximum error that is
due to scattered light. From our spectra from NARVAL and ESPaDOnS  we measured
abundances that are 0.03 and 0.04~dex too low.  We have therefore added 0.03~dex
to the abundances of all stars.

\section{Spectral analysis of \kepler\ targets}

In the initial spectral analysis of the \kepler\ stars we used fixed starting
parameters, namely $T_{\rm eff} = 6000$\,K, $\log g = 4.0$ and $v_{\rm micro} =
1.0$\,\kms.  The \vsini\ was adjusted manually by comparing synthetic spectra to
isolated lines in the range 6000 -- 6150\,\AA. { Values for macroturbulence
were taken from the calibration of \citet{bruntt23}.} The final \vsini\ was
adjusted later, once the other atmospheric parameters had been improved. VWA was
run in an automatic mode, where the atmospheric parameters were iteratively
adjusted. This was done by minimizing the correlations of the \feone\ abundance 
with both equivalent width (EW) and excitation potential (EP), as described by
\cite{bruntt23,bruntt10-corot}. This was achieved by adjusting \teff\ and
\vmicro.  Also, we demanded that \feone\ and \fetwo\ agreed,  which is dependent
on both \teff\ and \logg. This iterative approach adopted  non-local
thermodynamical equilibrium (NLTE) corrections to \feone\ using interpolations
and extrapolations in the figures in \cite{holm96}.  This is exactly the same
approach used in previous analyses with VWA, but in the next section we describe
an important change in the analysis, utilizing the asteroseismic results.

\begin{table*}
\begin{centering}
 \begin{minipage}{140mm}
  \caption{Observed targets and their properties. 
The KIC parameters are given in columns 7--9, spectroscopic parameters in columns 9--10, and
the final values of \teff, \logg, [Fe/H], and \vmicro\ are given in columns 11--14, labelled ``asteroseismic \logg''.
When the asteroseismic \logg\ is used the uncertainties are: 60~K for
\teff, 0.03~dex for \logg, 0.06~\kms\ for \vmicro\ and 0.06~dex for [Fe/H]. 
For the spectroscopic values, the uncertainties are 0.08~dex for \logg\ and 70~K for \teff.
Note that we only list the the ``asteroseismic'' values of \vsini\ and $v_{\rm micro}$,
since the ``spectroscopic'' are almost identical.}
\label{tab:master}
\setlength{\tabcolsep}{2pt} 
  \begin{tabular}{rrrr rrrr rrrr rrrr}
  \hline
\multicolumn{6}{c}{} & 
\multicolumn{3}{c}{``KIC param.''} &
\multicolumn{2}{c|}{``Spect.~\logg''} &
\multicolumn{4}{c|}{``Asteroseismic \logg''}   &   \\ \hline
   KIC-ID& HIP   &  HD   & $V_{\rm T}$ & $V_{\rm T}-K_{\rm S}$ & $E(B-V)$ & \teff & \logg & \feh & \teff & \logg & \teff & \logg & \feh & $v_{\rm micro}$ & \vsini \\
         &       &       &             &                     &            & [K]   &       &      & [K]   &       & [K]   &       &      & [\kms]          & [\kms] \\ \hline
  1430163&       &       &  9.627 &1.098& $              $ &  6910  & 3.66 & $-0.05$ & 6574 &  4.27 & 6520 & 4.22 & $ -0.11$ &   1.64 & 11.0 \\
  1435467&       &       &  9.017 &1.299& $  0.01 \pm0.02$ &  6570  & 4.02 & $-0.25$ & 6345 &  4.24 & 6264 & 4.09 & $ -0.01$ &   1.45 & 10.0 \\
  2837475&       & 179260&  8.547 &1.083& $              $ &  6444  & 4.07 & $-0.06$ & 6740 &  4.51 & 6700 & 4.16 & $ -0.02$ &   2.35 & 23.5 \\
  3424541&       &       &  9.895 &1.281& $              $ &  6207  & 3.58 & $-0.12$ & 6180 &  3.50 & 6080 & 3.82 & $  0.01$ &   1.39 & 29.8 \\
  3427720&       &       &  9.280 &1.454& $  0.00 \pm0.04$ &  5780  & 4.44 & $-0.37$ & 6070 &  4.51 & 6040 & 4.38 & $ -0.03$ &   1.16 &  4.0 \\
  3456181&       &       &  9.776 &1.297& $  0.03 \pm0.02$ &  6361  & 3.54 & $-0.66$ & 6344 &  4.05 & 6270 & 3.93 & $ -0.19$ &   1.53 &  7.8 \\
  3632418&  94112& 179070&  8.310 &1.365& $  0.00 \pm0.04$ &        &      & $     $ & 6235 &  4.14 & 6190 & 4.00 & $ -0.16$ &   1.42 &  6.5 \\
  3656476&       &       &  9.643 &1.635& $  0.00 \pm0.04$ &  5424  & 4.47 & $-0.60$ & 5720 &  4.31 & 5710 & 4.23 & $  0.34$ &   1.02 &  2.0 \\
  3733735&  94071& 178971&  8.409 &1.035& $  0.02 \pm0.02$ &  6442  & 4.03 & $-0.62$ & 6715 &  4.45 & 6715 & 4.26 & $ -0.04$ &   1.99 & 16.8 \\
  4586099&       &       &  9.321 &1.341& $  0.03 \pm0.02$ &  6037  & 3.93 & $-0.77$ & 6296 &  4.05 & 6296 & 4.02 & $ -0.17$ &   1.50 &  6.1 \\
  4638884&       &       &  9.934 &1.182& $              $ &  6367  & 4.07 & $-0.31$ & 6380 &  4.01 & 6375 & 4.03 & $ -0.03$ &   1.72 &  7.8 \\
  4914923&  94734&       &  9.571 &1.636& $              $ &        &      & $     $ & 5880 &  4.30 & 5905 & 4.21 & $  0.17$ &   1.19 &  3.6 \\
  5021689&       &       &  9.588 &1.392& $  0.01 \pm0.02$ &  5997  & 4.29 & $-0.32$ & 6150 &  4.11 & 6168 & 4.02 & $ -0.08$ &   1.39 &  9.8 \\
  5184732&       &       &  8.386 &1.565& $              $ &  5599  & 4.31 & $-0.09$ & 5865 &  4.38 & 5840 & 4.26 & $  0.38$ &   1.13 &  3.3 \\
  5371516&  96528& 185457&  8.483 &1.281& $  0.02 \pm0.02$ &  6050  & 4.17 & $-0.11$ & 6408 &  4.34 & 6408 & 3.98 & $  0.17$ &   1.61 & 15.2 \\
  5450445&       &       &  9.880 &1.378& $              $ &  5965  & 4.20 & $-0.30$ & 6110 &  4.15 & 6112 & 3.98 & $  0.05$ &   1.36 &  8.9 \\
  5512589&       &       & 10.229 &1.666& $              $ &  5554  & 4.39 & $-0.32$ & 5780 &  4.16 & 5750 & 4.06 & $  0.06$ &   1.11 &  3.5 \\
  5596656&       &       &  9.867 &2.244& $  0.00 \pm0.04$ &  4942  & 4.29 & $-1.43$ & 5014 &  3.48 & 5094 & 3.35 & $ -0.46$ &   0.83 &  4.0 \\
  5773345&       &       &  9.331 &1.385& $              $ &  6009  & 4.27 & $ 0.08$ & 6184 &  4.08 & 6130 & 4.00 & $  0.21$ &   1.46 &  6.6 \\
  5774694&  93657& 177780&  8.439 &1.508& $              $ &        &      & $     $ & 5878 &  4.66 & 5875 & 4.47 & $  0.07$ &   1.02 &  5.2 \\
  5939450&  92771& 175576&  7.384 &1.284& $  0.01 \pm0.02$ &        &      & $     $ & 6200 &  3.61 & 6282 & 3.70 & $ -0.18$ &   1.56 &  8.3 \\
  5955122&       &       &  9.446 &1.527& $  0.02 \pm0.02$ &  5747  & 4.39 & $-0.21$ & 5865 &  3.88 & 5837 & 3.87 & $ -0.17$ &   1.22 &  6.5 \\
  6106415&  93427& 177153&  7.275 &1.446& $  0.00 \pm0.04$ &        &      & $     $ & 6055 &  4.40 & 5990 & 4.31 & $ -0.09$ &   1.15 &  4.0 \\
  6116048&       &       &  8.540 &1.419& $  0.01 \pm0.02$ &  5863  & 4.32 & $-0.07$ & 5990 &  4.32 & 5935 & 4.28 & $ -0.24$ &   1.02 &  4.0 \\
  6225718&  97527& 187637&  7.580 &1.297& $              $ &        &      & $     $ & 6250 &  4.46 & 6230 & 4.32 & $ -0.17$ &   1.38 &  5.0 \\
  6442183&       & 183159&  8.661 &1.678& $  0.02 \pm0.02$ &        &      & $     $ & 5740 &  4.16 & 5760 & 4.03 & $ -0.11$ &   1.06 &  4.5 \\
  6508366&       &       &  9.078 &1.277& $  0.03 \pm0.02$ &  6271  & 4.14 & $-0.21$ & 6310 &  3.95 & 6354 & 3.94 & $ -0.08$ &   1.52 & 18.2 \\
  6603624&       &       &  9.282 &1.716& $  0.01 \pm0.02$ &  5416  & 4.38 & $-0.06$ & 5640 &  4.47 & 5625 & 4.32 & $  0.28$ &   1.05 &  3.0 \\
  6679371&       &       &  8.792 &1.192& $              $ &  6313  & 4.07 & $-0.17$ & 6390 &  3.91 & 6260 & 3.92 & $ -0.13$ &   1.62 & 19.2 \\
  6933899&       &       &  9.818 &1.647& $  0.01 \pm0.02$ &  5616  & 4.25 & $-0.09$ & 5870 &  4.07 & 5860 & 4.09 & $  0.02$ &   1.15 &  3.5 \\
  7103006&       &       &  8.919 &1.217& $              $ &  6203  & 4.26 & $-0.01$ & 6394 &  4.18 & 6394 & 4.01 & $  0.05$ &   1.58 & 13.2 \\
  7206837&       &       &  9.908 &1.333& $              $ &  6100  & 4.15 & $-0.23$ & 6304 &  4.21 & 6304 & 4.17 & $  0.14$ &   1.29 & 10.1 \\
  7282890&       &       &  9.155 &1.214& $  0.03 \pm0.02$ &  6072  & 4.05 & $-0.44$ & 6410 &  4.11 & 6384 & 3.88 & $  0.02$ &   1.74 & 22.6 \\
  7510397&  93511& 177412&  7.922 &1.378& $  0.01 \pm0.02$ &  5972  & 4.13 & $-0.35$ & 6130 &  4.07 & 6110 & 4.01 & $ -0.23$ &   1.35 &  5.4 \\
  7529180&       &       &  8.542 &1.094& $              $ &  6367  & 3.52 & $-0.43$ & 6790 &  4.70 & 6700 & 4.23 & $ -0.02$ &   1.81 & 33.6 \\
  7662428&       &       &  9.797 &1.888& $              $ &  6088  & 4.18 & $-0.04$ & 6360 &  4.35 & 6360 & 4.27 & $  0.22$ &   1.46 & 15.7 \\
  7668623&       &       &  9.441 &1.286& $              $ &  6013  & 4.11 & $-0.29$ & 6270 &  4.01 & 6270 & 3.87 & $ -0.02$ &   1.43 &  9.6 \\
  7680114&       &       & 10.227 &1.554& $  0.02 \pm0.02$ &  5578  & 4.63 & $-0.33$ & 5825 &  4.25 & 5855 & 4.18 & $  0.11$ &   1.10 &  3.0 \\
  7747078&  94918&       &  9.593 &1.617& $  0.03 \pm0.02$ &        &      & $     $ & 5910 &  3.98 & 5840 & 3.91 & $ -0.26$ &   1.22 &  6.0 \\
  7799349&       &       &  9.782 &2.264& $              $ &  4920  & 3.47 & $ 0.18$ & 5065 &  3.71 & 5115 & 3.67 & $  0.41$ &   1.21 &  1.0 \\
  7800289&       &       &  9.599 &1.218& $              $ &  6208  & 4.21 & $-0.35$ & 6389 &  3.79 & 6354 & 3.71 & $ -0.23$ &   1.66 & 21.6 \\
  7871531&       &       &  9.512 &1.996& $  0.00 \pm0.04$ &  5153  & 4.16 & $-0.44$ & 5400 &  4.43 & 5400 & 4.49 & $ -0.24$ &   0.71 &  2.9 \\
  7940546&  92615& 175226&  7.477 &1.303& $              $ &  5987  & 4.17 & $-0.14$ & 6240 &  4.11 & 6264 & 3.99 & $ -0.19$ &   1.56 &  9.7 \\
  7970740&       & 186306&  8.042 &1.957& $  0.00 \pm0.04$ &  5231  & 4.41 & $-0.18$ & 5290 &  4.57 & 5290 & 4.58 & $ -0.49$ &   0.68 &  3.0 \\
  7976303&       &       &  9.220 &1.490& $  0.01 \pm0.02$ &  6005  & 4.23 & $-0.59$ & 6095 &  4.03 & 6053 & 3.87 & $ -0.53$ &   1.37 &  5.5 \\
  8006161&  91949&       &  7.610 &1.940& $              $ &  5184  & 3.63 & $ 0.36$ & 5370 &  4.51 & 5390 & 4.49 & $  0.34$ &   1.07 &  2.5 \\
  8026226&       &       &  8.529 &1.300& $  0.02 \pm0.02$ &  5930  & 3.84 & $-0.85$ & 6260 &  3.80 & 6230 & 3.71 & $ -0.16$ &   1.58 & 10.6 \\
  8179536&       &       &  9.570 &1.292& $  0.02 \pm0.02$ &  6163  & 4.12 & $-0.20$ & 6369 &  4.39 & 6344 & 4.27 & $  0.01$ &   1.44 &  9.0 \\
  8228742&  95098&       &  9.564 &1.448& $  0.04 \pm0.02$ &  5858  & 3.97 & $-0.26$ & 6070 &  4.00 & 6042 & 4.02 & $ -0.14$ &   1.30 &  5.7 \\
  8360349&       & 181777&  8.719 &1.279& $  0.00 \pm0.04$ &  6044  & 4.21 & $-0.09$ & 6340 &  4.15 & 6340 & 3.81 & $  0.12$ &   1.61 & 17.9 \\
  8367710&       &       &  9.987 &1.238& $              $ &  6166  & 4.31 & $-0.01$ & 6465 &  4.17 & 6500 & 3.99 & $  0.16$ &   1.70 & 21.1 \\
  8394589&       &       &  9.648 &1.422& $              $ &  5938  & 3.82 & $-0.33$ & 6130 &  4.33 & 6114 & 4.32 & $ -0.36$ &   1.23 &  6.0 \\
  8524425&       &       &  9.995 &1.818& $  0.03 \pm0.02$ &  5449  & 3.92 & $-0.21$ & 5620 &  4.03 & 5634 & 3.98 & $  0.14$ &   1.23 &  2.3 \\
  8542853&       &       & 10.382 &2.579& $  0.00 \pm0.04$ &  5406  & 4.41 & $-0.15$ & 5560 &  4.67 & 5560 & 4.54 & $ -0.20$ &   0.80 &  3.5 \\
  8561221&       &       & 10.166 &2.076& $  0.03 \pm0.02$ &  4997  & 4.65 & $-0.77$ & 5300 &  3.76 & 5245 & 3.61 & $ -0.06$ &   1.01 &  3.5 \\
  8579578&       &       &  8.996 &1.228& $  0.03 \pm0.02$ &  6078  & 4.21 & $-0.22$ & 6470 &  4.27 & 6380 & 3.92 & $ -0.09$ &   1.92 & 24.7 \\

\hline
\end{tabular}
\end{minipage}
\end{centering}
\end{table*}

\begin{table*}
\addtocounter{table}{-1}
\begin{centering}
 \begin{minipage}{140mm}
  \caption{Observed targets and their properties (table continued)}
\setlength{\tabcolsep}{2pt} 
  \begin{tabular}{rrrr rrrr rrrr rrrr}
  \hline
\multicolumn{6}{c}{} & 
\multicolumn{3}{c}{``KIC param.''} &
\multicolumn{2}{c|}{``Spect.~\logg''} &
\multicolumn{4}{c|}{``Asteroseismic \logg''}   &   \\ \hline
   KIC-ID& HIP   &  HD   & $V_{\rm T}$ & $V_{\rm T}-K_{\rm S}$ & $E(B-V)$ & \teff & \logg & \feh & \teff & \logg & \teff & \logg & \feh & $v_{\rm micro}$ & \vsini \\
         &       &       &             &                     &            & [K]   &       &      & [K]   &       & [K]   &       &      & [\kms]          & [\kms] \\ \hline
  8694723&       &       &  8.977 &1.314& $  0.03 \pm0.02$ &  6101  & 4.15 & $-0.51$ & 6200 &  4.08 & 6120 & 4.10 & $ -0.59$ &   1.39 &  6.6 \\
  8702606&       &       &  9.595 &1.911& $  0.02 \pm0.02$ &  5308  & 4.60 & $-0.48$ & 5540 &  3.92 & 5540 & 3.76 & $ -0.09$ &   1.08 &  4.0 \\
  8738809&       &       & 10.133 &1.452& $              $ &  5848  & 4.38 & $-0.15$ & 6090 &  4.02 & 6090 & 3.90 & $  0.11$ &   1.33 &  4.9 \\
  8751420&  95362& 182736&  7.103 &2.075& $  0.00 \pm0.02$ &        &      & $     $ & 5294 &  3.80 & 5264 & 3.70 & $ -0.15$ &   1.00 &  3.5 \\
  8760414&       &       &  9.752 &1.579& $  0.02 \pm0.02$ &        &      & $     $ & 5795 &  4.25 & 5787 & 4.33 & $ -1.14$ &   1.03 &  3.0 \\
  8938364&       &       & 10.287 &1.651& $  0.00 \pm0.04$ &  5741  & 4.34 & $-0.38$ & 5620 &  4.17 & 5630 & 4.16 & $ -0.20$ &   1.00 &  3.8 \\
  9098294&       &       &  9.980 &1.616& $  0.02 \pm0.02$ &  5701  & 4.44 & $-0.73$ & 5770 &  4.26 & 5840 & 4.30 & $ -0.13$ &   1.01 &  4.0 \\
  9139151&  92961&       &  9.351 &1.399& $  0.00 \pm0.04$ &  5911  & 4.32 & $-0.15$ & 6105 &  4.54 & 6125 & 4.38 & $  0.11$ &   1.22 &  6.0 \\
  9139163&  92962& 176071&  8.386 &1.155& $  0.00 \pm0.04$ &  6220  & 4.28 & $-0.13$ & 6375 &  4.27 & 6400 & 4.18 & $  0.15$ &   1.31 &  4.0 \\
  9206432&  93607&       &  9.152 &1.085& $  0.01 \pm0.02$ &  6301  & 4.27 & $-0.06$ & 6608 &  4.45 & 6608 & 4.23 & $  0.23$ &   1.70 &  6.7 \\
  9226926&       &       &  8.741 &1.032& $  0.02 \pm0.02$ &  6735  & 3.57 & $-1.01$ & 6892 &  4.49 & 6892 & 4.14 & $ -0.23$ &   2.04 & 31.1 \\
  9812850&       &       &  9.576 &1.277& $  0.01 \pm0.02$ &  6169  & 4.21 & $-0.29$ & 6330 &  4.16 & 6325 & 4.05 & $ -0.16$ &   1.61 & 12.9 \\
  9908400&       &       &  9.157 &1.332& $              $ &  5823  & 4.18 & $ 0.04$ & 6620 &  4.45 & 6400 & 3.74 & $  0.34$ &   1.65 & 26.1 \\
  9955598&       &       &  9.717 &1.949& $  0.00 \pm0.04$ &  5333  & 3.71 & $ 0.19$ & 5395 &  4.52 & 5410 & 4.48 & $  0.11$ &   0.87 &  2.0 \\
 10016239&       &       &  9.871 &1.130& $              $ &  6232  & 4.29 & $-0.58$ & 6340 &  4.33 & 6340 & 4.31 & $ -0.05$ &   1.42 & 14.7 \\
 10018963&       &       &  8.832 &1.379& $  0.00 \pm0.04$ &  5977  & 4.16 & $-0.47$ & 6130 &  3.98 & 6020 & 3.95 & $ -0.35$ &   1.32 &  5.5 \\
 10068307&  94675& 180867&  8.299 &1.355& $  0.00 \pm0.04$ &  6052  & 4.13 & $-0.21$ & 6140 &  4.00 & 6114 & 3.93 & $ -0.22$ &   1.38 &  6.5 \\
 10124866&  93108& 176465&  8.674 &2.334& $  0.00 \pm0.04$ &  5813  & 4.24 & $-0.14$ & 5850 &  4.55 & 5755 & 4.48 & $ -0.30$ &   0.82 &  3.0 \\
 10162436&  97992&       &  8.747 &1.389& $  0.01 \pm0.02$ &  6095  & 4.18 & $-0.23$ & 6180 &  4.02 & 6200 & 3.95 & $ -0.08$ &   1.44 &  6.5 \\
 10355856&       &       &  9.294 &1.237& $  0.00 \pm0.04$ &  6288  & 4.06 & $-0.53$ & 6395 &  4.12 & 6350 & 4.08 & $ -0.19$ &   1.55 &  7.2 \\
 10454113&  92983&       &  8.744 &1.453& $  0.00 \pm0.04$ &  5972  & 4.30 & $-0.11$ & 6120 &  4.32 & 6120 & 4.31 & $ -0.06$ &   1.21 &  5.5 \\
 10462940&       &       &  9.836 &1.368& $  0.00 \pm0.04$ &  6070  & 4.27 & $-0.19$ & 6174 &  4.46 & 6154 & 4.32 & $  0.10$ &   1.29 &  4.3 \\
 10516096&       &       &  9.611 &1.482& $  0.02 \pm0.02$ &  5814  & 4.19 & $-0.31$ & 5940 &  4.21 & 5940 & 4.18 & $ -0.06$ &   1.12 &  4.2 \\
 10644253&       &       &  9.324 &1.450& $  0.00 \pm0.04$ &  5831  & 4.28 & $-0.15$ & 6105 &  4.59 & 6030 & 4.40 & $  0.12$ &   1.14 &  3.8 \\
 10709834&       &       &  9.859 &1.071& $  0.00 \pm0.04$ &  6502  & 3.99 & $-0.23$ & 6570 &  4.17 & 6508 & 4.09 & $ -0.08$ &   1.66 & 10.6 \\
 10923629&       &       &  9.936 &1.300& $  0.03 \pm0.02$ &  5982  & 4.26 & $-0.00$ & 6214 &  3.95 & 6214 & 3.82 & $  0.16$ &   1.50 & 11.4 \\
 10963065&       &       &  8.893 &1.407& $  0.01 \pm0.02$ &  6043  & 4.16 & $-0.25$ & 6090 &  4.31 & 6060 & 4.29 & $ -0.20$ &   1.06 &  5.4 \\
 11026764&       &       &  9.783 &1.921& $  0.00 \pm0.04$ &  5502  & 3.90 & $-0.25$ & 5670 &  4.01 & 5682 & 3.88 & $  0.05$ &   1.17 &  4.8 \\
 11081729&       &       &  9.061 &1.088& $  0.02 \pm0.02$ &  6359  & 3.98 & $-0.10$ & 6630 &  4.41 & 6630 & 4.25 & $ -0.12$ &   1.72 & 20.0 \\
 11137075&       &       & 11.034 &1.760& $  0.01 \pm0.02$ &  5425  & 4.17 & $ 0.00$ & 5590 &  4.30 & 5590 & 4.01 & $ -0.06$ &   0.96 &  1.0 \\
 11244118&       &       &  9.937 &1.643& $              $ &  5507  & 4.50 & $ 0.14$ & 5735 &  4.23 & 5745 & 4.09 & $  0.35$ &   1.16 &  3.0 \\
 11253226&  97071& 186700&  8.514 &1.055& $  0.00 \pm0.02$ &  6468  & 4.18 & $-0.34$ & 6605 &  4.21 & 6605 & 4.16 & $ -0.08$ &   1.73 & 15.1 \\
 11414712&       &       &  8.785 &1.752& $  0.00 \pm0.04$ &  5388  & 3.80 & $-0.66$ & 5620 &  3.84 & 5635 & 3.80 & $ -0.05$ &   1.18 &  4.4 \\
 11498538&  93951& 178874&  7.420 &1.172& $  0.03 \pm0.02$ &  6287  & 4.04 & $-0.30$ & 6496 &  3.83 & 6496 & 3.76 & $  0.01$ &   1.77 & 38.4 \\
 11717120&       &       &  9.586 &2.180& $  0.00 \pm0.04$ &  4980  & 3.44 & $-0.45$ & 5105 &  3.80 & 5150 & 3.68 & $ -0.30$ &   0.98 &  1.0 \\
 12009504&       &       &  9.380 &1.311& $  0.00 \pm0.02$ &  6056  & 3.70 & $ 0.13$ & 6125 &  4.23 & 6065 & 4.21 & $ -0.09$ &   1.13 &  8.4 \\
 12258514&  95568& 183298&  8.204 &1.446& $  0.00 \pm0.04$ &  5808  & 4.30 & $ 0.09$ & 6025 &  4.28 & 5990 & 4.11 & $  0.04$ &   1.22 &  3.5 \\
\hline
\end{tabular}
\end{minipage}
\end{centering}
\end{table*}

\begin{figure} 
\begin{center} 
 \includegraphics[width=8.8cm]{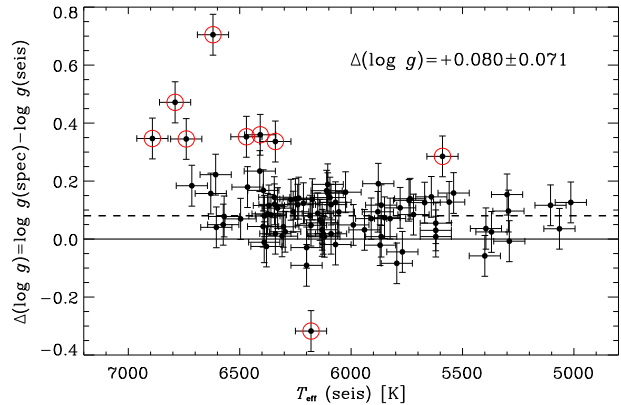}
 \caption{The difference in \logg\ determined from spectroscopic 
and asteroseismic methods. The spectroscopic value is systematically higher by
about 0.08 dex with RMS scatter of 0.07 dex (dashed line indicates the mean difference). 
A few apparent outliers have been marked by open symbols and discussed in 
Sect.~\ref{sec:logg}.
\label{fig:logg}}
\end{center} 
\end{figure} 

\subsection{Adopting asteroseismic \logg\ values\label{sec:logg}}

The determination of stellar \logg\ values from spectroscopy alone is
notoriously unreliable. Although internal errors of less than 0.1~dex are often
reported from high-quality data, realistic uncertainties are more likely of the
order $\pm$0.2~dex \citep{2005MSAIS...8..130S}. Asteroseismically determined
\logg\ values are significantly more precise. This is due to the measurement of
stellar oscillations, whose gross properties (large separation and frequency of
maximum amplitude) depend on the mass, radius and effective temperature via
well-established scaling relations
\citep{1991ApJ...371..396B,1995A&A...293...87K}. The surface gravity scales as
$M / R^2$ and is therefore very well determined from asteroseismology
\citep{2010ApJ...713L.164C}. This allows us to further constrain the other
spectroscopically determined parameters. For example, changing \teff\ affects
how \feone\ depends on EP and both \logg\ and \teff\ affect the \feone\ /
\fetwo\ ionization balance, which can lead to a strong correlation of the output
parameters of \teff\ and \logg.  This was seen in the analysis of the solar-like
\kepler\ target KIC~11026764 reported by \cite{metcalfe10}, where the \logg\
uncertainty was reduced from 0.3~dex to 0.02~dex using asteroseismology.

In the present work, we have adopted the same approach of fixing the \logg\ to
the asteroseismically determined values. The asteroseismic {\logg} were
estimated with the grid-based Yale-Birmingham pipeline \citep{basu10,gai11},
which used as input the seismic parameters $\Delta\nu$ and $\nu_{\rm max}$, and
the spectroscopically determined $T_{\rm eff}$ and [Fe/H], while adopting the
Yonsei-Yale grid of evolution models (see \citealt{basu10} for details).  The
seismic parameters came from independent analyses of the \textsl{Kepler}
lightcurves performed by five different teams (details are given by
\citealt{2011ApJ...738L..28V,2011MNRAS.415.3539V}).

We found that the difference between different model grids mentioned in
\cite{basu10} resulted in changes in the estimated \logg\ of less than 0.02 dex
in 90 percent of the stars and less than 0.05 dex for all stars. These are
smaller than the typical uncertainties in \logg\ when determined from the
spectrum alone, which are typically 0.08~dex for the slowly rotating stars in
the sample, and slightly larger for the moderately fast rotators.

The determined stellar parameters are given in Table~\ref{tab:master}, both
using the ``asteroseismic \logg'' in columns 12--15 and the classical
``spectroscopic method'' (e.g.\ as described in \citealt{bruntt23}) in columns
10--11.  For completion we also list the KIC parameters in columns 7--9.  The
differences in determined parameters between the two methods are generally
small, with mean differences and RMS values as follows $\Delta \log g = +0.08
\pm 0.07$\,dex, $\Delta T_{\rm eff} = +10 \pm 38$\,K, $\Delta {\rm [Fe/H]} =
-0.00 \pm 0.03$\,dex, and $\Delta v_{\rm micro} = +0.02 \pm 0.09$\,\kms
(calculated in the sense ``classical'' {\em minus} seismic).  Considering that
the error on the mean values is $\sqrt {N} \simeq 9$ times smaller than the RMS
values given here, only the difference in \logg\ is significant (10\,$\sigma$),
while the differences in \teff\ and \vmicro\ are at the $2\,\sigma$ level.

To further investigate the significant discrepancy in \logg, we show in
Fig.~\ref{fig:logg} the difference between the \logg\ found from the two
different methods.  We see that the difference does not depend on spectral
type.  We cannot offer an explanation for the systematic offset, but the quite
low RMS scatter of 0.07 dex indicates that the two methods are internally
consistent. Further analyses are needed to understand these differences.  Use of
independent estimates, provided by analysing binary stars that also show
pulsating solar-type components, would be highly desirable in this context and
might be possible in future space missions like {\em PLATO} \citep{catala09}. We
note that from Fig.~\ref{fig:logg}, the ``classical'' method appears to be able
to determine \logg\ with an { accuracy} of about 0.10 dex for solar-type stars.

The stars with large offsets are marked by circles in Fig.~\ref{fig:logg} and
have KIC-IDs:   
  2837475,
  3424541, 
  5371516, 
  7529180, 
  8360349, 
  8579578, 
  9226926, 
  9908400, and 
 11137075. The last star has parameters close to the Sun but the
others have \teff\ in the range 6080 -- 6890~K, \logg\ from 3.7 to 4.2, 
and \vsini\ between 15 and 34\,\kms. One star (3424541) has only 2 usable
\fetwo\ lines, while the others have at least 6 and up to 17. 
We do not find an immediate explanation for these apparent outliers in \logg.

In the remainder of this paper we use the stellar parameters from
using the asteroseismic \logg, and we have labelled the axes in the 
figures \teff\,(seis) and \logg\,(seis). 



\begin{figure} 
\begin{center} 
 \includegraphics[width=8.8cm]{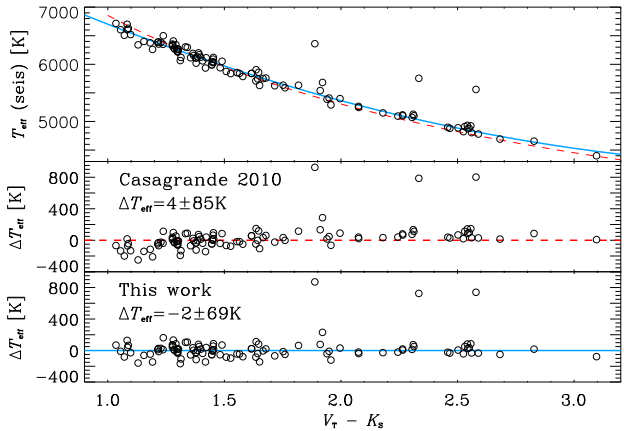}
 \caption{The upper panel shows \teff\,(seis) vs.\ $V_{\rm T} - K_{\rm S}$ 
for stars with at least 100 \feone\ lines. 
The three stars that lie above the general distribution are close binary systems.
The solid line is the fit from Eq.~\ref{eq:teff} and
the dashed line is the fit from \citet{casagrande10}.
The middle and lower panels show the difference in
\teff\ for the two calibrations.
\label{fig:teff}}
\end{center} 
\end{figure} 

\subsection{Photometric calibration of \teff}

We have used the adopted spectroscopic parameters to make a calibration of
\teff\ vs.\ the photometric index $V_{\rm T} - K_{\rm S}$, where $V_{\rm T}$ is the TYCHO $V$
magnitude and $K_{\rm S}$ is the 2MASS magnitude. All stars in our sample have
measured values of these magnitudes. We adopted this index because it is known
to be a very good indicator of \teff. For example, \cite{casagrande10} found
this index to have one of the lowest RMS residuals among the many indices they
used for calibrating \teff.

In Fig.~\ref{fig:teff} we plot \teff\,(seis) vs.\ the colour index 
$V_{\rm T} - K_{\rm S}$. The dashed line is the calibration of this index taken from
\cite{casagrande10} and the residuals from the fit are shown in the middle
panel. The offset of 4~K is negligible compared to the RMS scatter of 85~K. The
solid line in the top panel is a second order fit and the residuals are shown in
the bottom panel, with a slightly lower RMS scatter of 69~K. The three obvious
outliers (KIC 7662428, 8542853, and 10124866), which are all close binary
systems with near-equal components, have been excluded  from the calculations of
the fit and the RMS values. The calibration is
\begin{equation}
T_{\rm eff}/{\rm K} = 8545 - 2090 \, (V_{\rm T}-K_{\rm S}) + 250 \, (V_{\rm T}-K_{\rm S})^2.
\label{eq:teff}
\end{equation}
This new empirical calibration is valid for both
main sequence and sub-giant stars ($\log g > 2.5$) with $V_{\rm T}-K_{\rm S}$ in the range
$1$ -- $3$. The 1-$\sigma$ uncertainty is 70~K, as measured from the RMS scatter.

To estimate the influence of interstellar reddening, we measured the strength of
the Na~D doublet at $5990 + 5995$\,\AA. In 70 stars, interstellar lines
were present and their equivalent widths were measured. The strength has been
calibrated by \cite{munari97} to give $E(B-V)$ with an uncertainty of 0.05 mag.
The values we determined are given in Table~\ref{tab:master}. We find that most
stars with a detectable interstellar Na~D doublet have almost negligible
reddening, with values in the range  from 0.00 to 0.06 mag. The two stars with
the highest values of $E(B-V)$ are KIC 3430868 and 8491147, which have values of
0.043 and 0.058. Since the calibration has an uncertainty of 
0.05 mag, we consider the interstellar reddening to be negligible for most of the
targets.

\subsection{A new calibration of microturbulence}

\begin{figure} 
\begin{center} 
 \includegraphics[width=8.8cm]{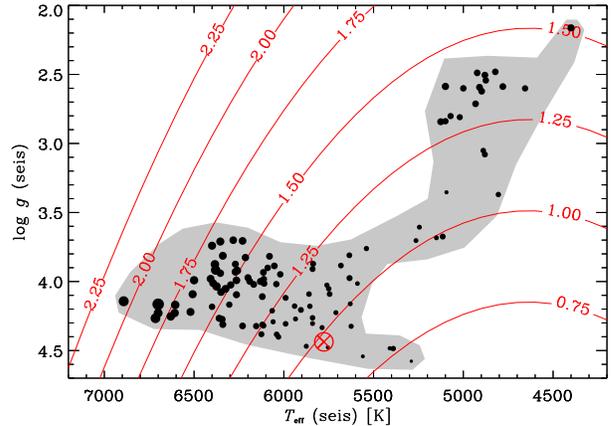}
 \caption{The diagram of \teff\,(seis) vs.\ asteroseismic \logg\ for 
\kepler\ targets.
The size of the points scale with the microturbulence. The contours show the
calibration of microturbulence and the gray area marks the region where 
it is valid.
The location of the Sun is marked with a cross and circle.
Stars with \logg\ below 3.5 are taken from Thygesen et al. (in preparation).
\label{fig:tg}}
\end{center} 
\end{figure} 

In the final determination of the stellar parameters, \logg\ was held
fixed at the asteroseismically determined value,
and only \teff\ and the microturbulence were allowed to vary. 
The locations of the stars in the \teff\ vs.\ \logg\ 
diagram are shown in Fig.~\ref{fig:tg}. 
We made a set of calibrations of the microturbulence using different
combinations of \teff\ and \logg\ and found that the following gave 
the lowest residuals:
\begin{eqnarray}
v_{\rm micro}/{\rm km\,s}^{-1} & = & 1.095 +\, 5.44\times 10^{-4} \, (T_{\rm eff} - 5700) \nonumber \\
              &   & +\, 2.56\times 10^{-7} \, (T_{\rm eff} - 5700)^2  \nonumber \\
              &   & -\, 0.378\, (\log g - 4.0)\, .
\end{eqnarray}
The RMS scatter of the residuals is 0.095\,\kms, reducing to 0.057\,\kms\ when a handful of
3-$\sigma$ outliers are removed.
A fit without the quadratic term in \teff\ gives 20\% higher residuals.
To be conservative, we adopt an uncertainty on this calibration of 0.1 km/s.
When using the above calibration, one must include 
the uncertainties in \teff\ and \logg. 
For typical uncertainties in \teff\ of 100\,K and 0.1 dex
in \logg, one finds a total uncertainty in \vmicro\ of 0.13 km/s.
This new calibration is in agreement with that presented by \cite{bruntt23},
but has a wider applicability due to the increased sample size.
The region of applicability is marked by the grey area in Fig.~\ref{fig:tg},
i.e.\ roughly 5300 -- 6900~K for main-sequence stars and 4500 -- 5200~K for the
evolved stars ($\log g$ from 2.1 to 3.5).

\begin{figure} 
\begin{center} 
 \includegraphics[width=8.8cm]{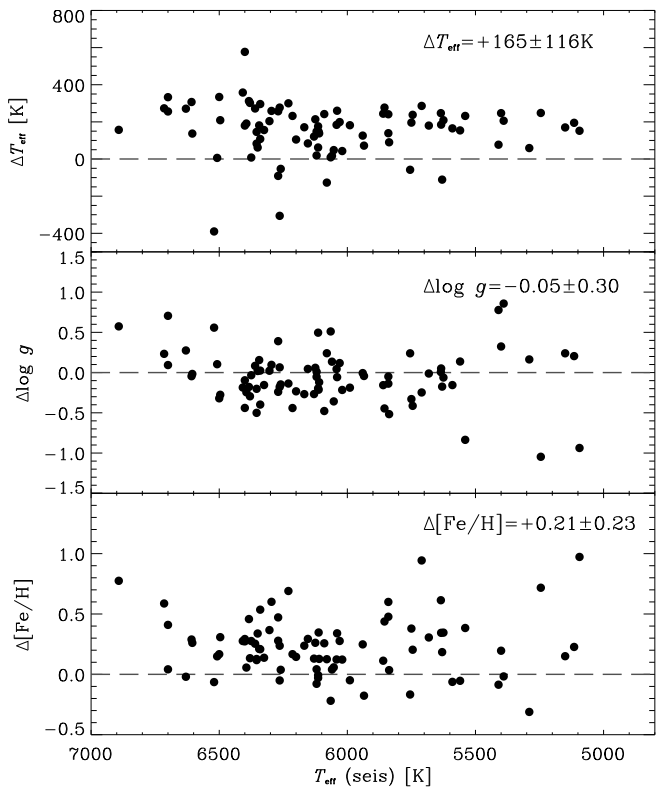}
 \caption{Differences between our adopted spectroscopic parameters
(\logg\ from asteroseismology) and the KIC photometric parameters 
of \teff, \logg\ and [Fe/H]. 
The ``$\Delta$'' means spectroscopic {\em minus} KIC parameter.
The mean offset  and RMS are given in each panel.
\label{fig:kic}}
\end{center} 
\end{figure} 

\section{Comparison with the KIC}

Out of 93 stars in our sample, 83 stars have stellar parameters in KIC \citep{2011AJ....142..112B}.
In Fig.~\ref{fig:kic} we compare our adopted spectroscopic values, 
i.e.\ with \logg\ fixed at the asteroseismically determined value,
of \teff, \logg, and [Fe/H] with the KIC.

In each panel we give the mean offset and RMS scatter.  These values
were calculated using a robust method where 3-$\sigma$ outliers were
removed (3, 4, 2 outliers were removed, respectively).  It seems the
KIC parameters are quite reliable across the entire spectral range
from 5100 to 6900~K, although there is a significant offset in \teff\
and [Fe/H].  Considering the approach of converting the ground-based
photometry to fundamental parameters described by \cite{2011AJ....142..112B},
which includes the non-trivial task of removing sky extinction and
interstellar reddening, we verify that the quality of the KIC
catalogue is generally very good.  Compared to our spectroscopic
values, we find that \teff\ in KIC are 165\,K lower and [Fe/H] are 0.21\,dex
lower. 

There is a modest, negative offset between the asteroseismic \logg\ and the KIC
\logg\ values. This is consistent with the results presented in
\cite{2011ApJ...738L..28V}, once allowance is made for differences in the
samples of stars used. They compared asteroseismic \logg\ with the KIC \logg\ of
more than 500 \textsl{Kepler} dwarfs and subgiants. They found reasonable
agreement between the two sets of values for KIC $\log g < 4$, but at
progressively higher values the KIC \logg\ were found to increasingly
overestimate the asteroiseismic $\log g$. The average offset, over the entire
ensemble, was about 0.17\,dex.

Most of the stars in this paper fall in the region $\log g \ge 4$, and the
significant offset identified by \cite{2011ApJ...738L..28V} is present in those
stars in our sample. However, there are only a few stars in this paper with KIC
$\log g < 4$, and they all have quite sizeable negative differences with respect
to the asteroseismic $\log g$. In contrast, the sampling of stars in Verner et
al. having $\log g < 4$ was much more comprehensive and showed a mean offset
close to zero. The effect of the sparse lower \logg\ sample here is to reduce
the mean offset from $\approx 0.17$\,dex to the $\approx 0.05$\,dex seen in
Fig.~\ref{fig:kic}.

\section{Consideration of NLTE effect on \feone}

\begin{figure} 
\begin{center} 
 \includegraphics[width=8.8cm]{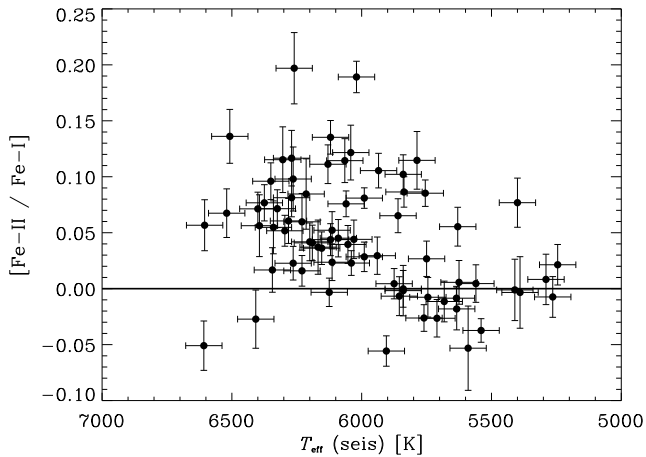}
  \caption{Difference in abundances determined from \fetwo\ and \feone\ in 79 unevolved
stars { using asteroseismic \logg\ values and
{\em without} applying any NLTE corrections.}
\label{fig:nlte}}
\end{center} 
\end{figure}

In Fig.~\ref{fig:nlte} we show the difference in abundance derived from Fe\ione\
and Fe\itwo\ lines, { {\em without} applying any NLTE corrections,} showing
evidence for a correlation with \teff. There is a large scatter, but most stars
with $T_{\rm eff} > 6000$\,K have a higher abundance from \fetwo. Indeed, it is
predicted by theoretical NLTE calculations for iron that, under the assumption
of LTE, the computed \feone\ lines will be too strong, hence the abundances
determined from \feone\ too small. The magnitude of this effect for A- and early
F-type stars was computed by \cite{holm96} based on a model of the Fe atom,
although recent calculations show that the effect is smaller
\citep{mashonkina09, mashonkina11}. However, our empirical results appear to be
in rough agreement with those of \cite{holm96}. As an example, for the hottest
stars in our sample (6500~K), $+0.07$~dex is the correction to be applied to
Fe\ione, while extrapolations from the Figures in \cite{holm96} indicate that
Fe\ione\  should be adjusted by $+0.05$~dex. 

{ Our analyses are strickly differential with respect to the Sun. Hence,
Fe\ione\ and Fe\itwo\ abundances are expected to agree for stars with similar
parameters. In fact, Fig.~\ref{fig:nlte} shows that this is the case for the
majority of these stars. Recall from Fig.~\ref{fig:logg} that there is an
apparent increase in the discrepancy in \logg\ for the hottest stars, which
could be the result of NLTE effects. However, the fact that the \logg\ offset
appears to be independent of \teff, except for these hotter stars, suggests that
this is not a result of NLTE effects alone, but could be due to other effects,
such as the shortcomings of mixing-length theory or 1D model atmospheres.} It is
beyond the scope of this paper to discuss this in more detail and we suggest
that a larger sample of \kepler\ stars, with both asteroseismic \logg\ values
and accurate spectroscopic \teff, is needed to place stronger constraints on
theoretical NLTE calculations.

\section{The abundance patterns}

\begin{figure} 
\begin{center} 
 \includegraphics[width=8.8cm]{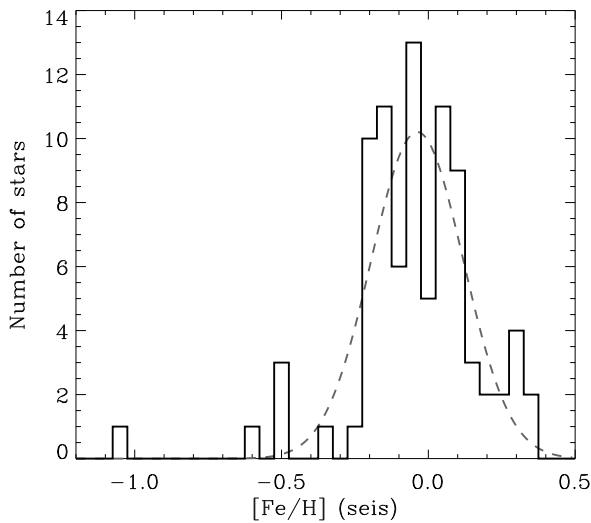}
 \caption{Histogram of the spectroscopically determined iron abundance.
The diagram shows [Fe/H] using only \fetwo\ lines as
determined for 85 stars with at least 10 \fetwo\ lines.
The dashed curve is a Gaussian fit.
{ The star with the lowest ${\rm [Fe/H]} = -1.14$ is KIC 8760414}.
\label{fig:feh}}
\end{center} 
\end{figure}

The distribution of metallicities is presented as a histogram in
Fig.~\ref{fig:feh}. Normally one would use \feone\ lines, since stellar spectra
typically have between 200 and 300 isolated lines for the slowly rotating stars in our
sample.  However, NLTE effects start to become important for \feone\ above
6000~K (Sect.~6). The \fetwo\ lines are less affected,  since singly-ionized
iron is the most common state of Fe, and so we used \fetwo\ as our primary
metallicity indicator. 

Figure~\ref{fig:feh} includes only stars with at least  10 \fetwo\ lines (85
stars). The peak value of the sample was computed  from a fit of a Gaussian
profile to the histogram (dashed bell-curve in Fig.~\ref{fig:feh}). 
The central value is $-0.04$ dex with a FWHM of $0.35$ dex.  We have used a bin size of 0.05
dex, but we found that changing this in the range 0.02 to 0.10 only changed  the
mean metallicity by 0.02 dex. Similarly, using \feone\ instead we found a mean
value of $-0.08$ dex. 

When performing the asteroseismic analysis of \kepler\ light curves, one often
does not have a reliable estimate of the metallicity. This is a problem when
comparing with theoretical evolution grids, especially for evolved stars
\citep{basu10}. Based on the distribution of metallicity in Fig.~\ref{fig:feh},
we recommend that for such cases one adopts  ${\rm [Fe/H]} = -0.07\pm0.20$. In
our sample, only seven stars  have metallicity below $-0.3$ and four stars are
above $+0.3$.  However, we note that this small sample of stars with ``extreme''
metallicity will be particularly interesting to study due to their different
internal opacity, which may affect the asteroseismic properties of the stars.

\begin{figure} 
\begin{center} 
 \includegraphics[width=8.8cm]{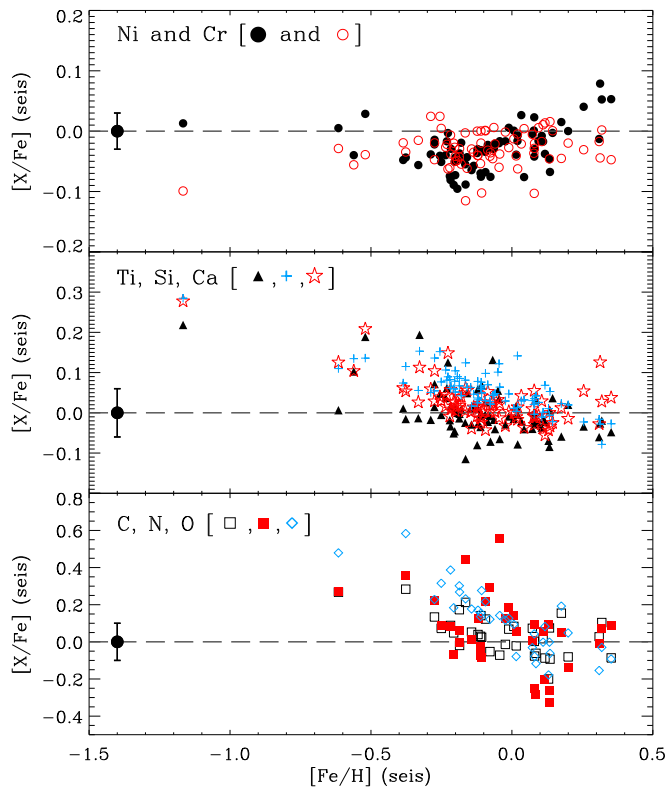}
 \caption{Abundance scaled with Fe versus the Fe abundance for selected elements.
The error bar around the solid circle at ${\rm [Fe/H]} = -1.4$ indicates the
approximate scatter of the points, which are 0.03, 0.06 and 0.10 dex for the three panels. 
The dashed horizontal line indicates the solar abundance. Note the
different scale in the lower panel.
\label{fig:elem}}
\end{center} 
\end{figure} 

\begin{table*}
 \centering
  \caption{Abundances relative to the Sun and the number of lines used for 13 elements. 
For each star the first and second line is for the neutral and singly ionized lines, respectively.
The adopted uncertainties are 0.07 dex for elements with more than five lines and 0.15 for other elements.
The complete table is available in the electronic version.
\label{tab:elem}}
\setlength{\tabcolsep}{1.7pt}
  \begin{tabular}{r rr rr rr rr rr rr rr rr rr rr rr rr rr}
\hline
    KIC~ID&  
   \multicolumn{2}{c}{Li} & \multicolumn{2}{c}{C} &\multicolumn{2}{c}{N}  &\multicolumn{2}{c}{O} &\multicolumn{2}{c}{Na}  &\multicolumn{2}{c}{Mg} &\multicolumn{2}{c}{Si} &
\multicolumn{2}{c}{Ca} &\multicolumn{2}{c}{Ti} &\multicolumn{2}{c}{V} &\multicolumn{2}{c}{Cr} &\multicolumn{2}{c}{Fe} &\multicolumn{2}{c}{Ni}\\ \hline
  1430163 &$     $&       &$ -0.09  $&$  11$&$ -0.13  $&$   2$&$ +0.03  $&$   4$&$ -0.14  $&$   5$&$ -0.08  $&$   2$&$ -0.18  $&$  18$&$ -0.13  $&$  20$&$ -0.17  $&$  10$&$ -0.10  $&$   1$&$ -0.17 $&$  18$&$ -0.14  $&$ 233$&$ -0.21  $&$  43$\\
          &$     $&       &$        $&$    $&$        $&$    $&$        $&$    $&$        $&$    $&$        $&$    $&$ +0.04  $&$   1$&$        $&$    $&$ -0.13  $&$  22$&$        $&$    $&$ -0.16 $&$  13$&$ -0.08  $&$  26$&$        $&$    $\\
  1435467 &$     $&       &$ -0.11  $&$  10$&$ +0.51  $&$   3$&$ +0.10  $&$   5$&$ +0.26  $&$   6$&$ -0.04  $&$   2$&$ -0.02  $&$  13$&$ +0.02  $&$   6$&$ -0.11  $&$  10$&$ +0.01  $&$   1$&$ -0.08 $&$  16$&$ -0.04  $&$ 188$&$ -0.06  $&$  21$\\
          &$     $&       &$        $&$    $&$        $&$    $&$        $&$    $&$        $&$    $&$ +0.07  $&$   2$&$ +0.16  $&$   1$&$        $&$    $&$ +0.04  $&$  19$&$        $&$    $&$ -0.04 $&$  11$&$ +0.05  $&$  19$&$        $&$    $\\
  2837475 &$     $&       &$ -0.05  $&$   7$&$ -0.31  $&$   3$&$ -0.06  $&$   1$&$ +0.05  $&$   2$&$ -0.06  $&$   2$&$ -0.02  $&$  14$&$ +0.05  $&$  10$&$ +0.01  $&$   7$&$        $&$   2$&$ -0.13 $&$   7$&$ -0.05  $&$ 133$&$ -0.04  $&$  30$\\
          &$     $&       &$        $&$    $&$        $&$    $&$        $&$    $&$        $&$    $&$ +0.00  $&$   2$&$ -0.09  $&$   1$&$        $&$    $&$ -0.13  $&$   8$&$        $&$    $&$ -0.20 $&$   5$&$ -0.05  $&$  12$&$        $&$    $\\
  3424541 &$+1.44$&     1 &$ +0.34  $&$   8$&$ +0.50  $&$   2$&$ +0.43  $&$   1$&$ +0.15  $&$   2$&$ -0.24  $&$   1$&$ +0.19  $&$  18$&$ +0.17  $&$   5$&$ +0.01  $&$   6$&$ -0.21  $&$   1$&$ -0.06 $&$  10$&$ -0.02  $&$  69$&$ +0.01  $&$  34$\\
          &$     $&       &$        $&$    $&$        $&$    $&$        $&$    $&$        $&$    $&$        $&$    $&$ +0.51  $&$   1$&$        $&$    $&$ +0.25  $&$   4$&$        $&$    $&$ +0.08 $&$   4$&$ +0.32  $&$   2$&$        $&$    $\\
  3427720 &$+1.53$&     1 &$ +0.05  $&$   7$&$ -0.07  $&$   1$&$ +0.06  $&$   2$&$ -0.05  $&$   4$&$ +0.04  $&$   2$&$ -0.04  $&$  28$&$ -0.02  $&$  21$&$ -0.08  $&$  26$&$ -0.05  $&$   5$&$ -0.06 $&$  16$&$ -0.06  $&$ 226$&$ -0.09  $&$  63$\\
          &$     $&       &$        $&$    $&$        $&$    $&$        $&$    $&$        $&$    $&$ +0.11  $&$   1$&$ -0.03  $&$   1$&$        $&$    $&$ +0.04  $&$  19$&$        $&$    $&$ -0.14 $&$   7$&$ -0.04  $&$  21$&$        $&$    $\\
  3430868 &$     $&       &$ +0.04  $&$   4$&$        $&$    $&$ +0.02  $&$   4$&$ +0.21  $&$   2$&$ +0.07  $&$   3$&$  0.08  $&$  21$&$ +0.11  $&$   9$&$ +0.02  $&$  29$&$ +0.09  $&$  11$&$ +0.02 $&$  11$&$ +0.06  $&$ 157$&$ -0.03  $&$  44$\\
          &$     $&       &$        $&$    $&$        $&$    $&$        $&$    $&$        $&$    $&$        $&$    $&$        $&$    $&$        $&$    $&$ +0.09  $&$  17$&$        $&$    $&$ +0.02 $&$   8$&$ +0.01  $&$  19$&$        $&$    $\\
  3456181 &$+1.59$&     1 &$ -0.09  $&$  10$&$ +0.01  $&$   1$&$ +0.02  $&$   5$&$ -0.20  $&$   4$&$ -0.15  $&$   2$&$ -0.23  $&$  24$&$ -0.14  $&$  21$&$ -0.24  $&$  19$&$ -0.21  $&$   3$&$ -0.28 $&$  20$&$ -0.22  $&$ 205$&$ -0.29  $&$  45$\\
          &$     $&       &$        $&$    $&$        $&$    $&$        $&$    $&$        $&$    $&$        $&$    $&$ +0.05  $&$   1$&$        $&$    $&$ -0.07  $&$  24$&$        $&$    $&$ -0.25 $&$   8$&$ -0.10  $&$  30$&$        $&$    $\\
  3632418 &$     $&       &$ -0.20  $&$  11$&$ -0.19  $&$   2$&$ +0.08  $&$   3$&$ -0.14  $&$   6$&$ -0.15  $&$   2$&$ -0.17  $&$  25$&$ -0.10  $&$  19$&$ -0.19  $&$  18$&$ -0.23  $&$   2$&$ -0.21 $&$  20$&$ -0.19  $&$ 291$&$ -0.22  $&$  54$\\
          &$     $&       &$        $&$    $&$        $&$    $&$        $&$    $&$        $&$    $&$ -0.10  $&$   1$&$ -0.11  $&$   1$&$        $&$    $&$ -0.09  $&$  21$&$        $&$    $&$ -0.22 $&$  11$&$ -0.14  $&$  29$&$        $&$    $\\
  3656476 &$     $&       &$ +0.34  $&$  12$&$ +0.30  $&$   2$&$ +0.16  $&$   5$&$ +0.34  $&$   4$&$ +0.42  $&$   2$&$ +0.28  $&$  36$&$ +0.29  $&$   9$&$ +0.25  $&$  22$&$ +0.29  $&$  17$&$ +0.29 $&$  10$&$ +0.31  $&$ 221$&$ +0.30  $&$  57$\\
          &$     $&       &$        $&$    $&$        $&$    $&$        $&$    $&$        $&$    $&$ +0.39  $&$   1$&$ +0.29  $&$   1$&$        $&$    $&$ +0.31  $&$  19$&$        $&$    $&$ +0.20 $&$   7$&$ +0.28  $&$  17$&$        $&$    $\\
  3733735 &$     $&       &$ -0.16  $&$  12$&$ -0.31  $&$   4$&$ +0.01  $&$   2$&$ -0.08  $&$   3$&$ -0.08  $&$   2$&$ -0.04  $&$  16$&$ -0.03  $&$  18$&$ +0.06  $&$  15$&$        $&$    $&$ -0.11 $&$  16$&$ -0.07  $&$ 147$&$ -0.10  $&$  35$\\
          &$     $&       &$        $&$    $&$        $&$    $&$        $&$    $&$        $&$    $&$ -0.01  $&$   2$&$ -0.01  $&$   1$&$        $&$    $&$ -0.12  $&$  20$&$        $&$    $&$ -0.15 $&$  12$&$ -0.07  $&$  18$&$        $&$    $\\
\hline
\end{tabular}
\end{table*}

In Fig.~\ref{fig:elem} we show the abundances of eight important elements versus
the metallicity. Ni and Cr have the most lines in the spectra after Fe. Ti, Si
and Ca also have many lines and are tracers of the $\alpha$ elements. C, N and O
have fewer lines but are important because they directly affect the fusion
processes in the core of the star. We only show abundances for stars with at
least 5 lines of each element,  except for N and O where 2 and 3 lines were
considered enough to compute the mean abundance. Also, only stars with \vsini\
below 25 \kms\ are included in the plots.  Around 92 and 95 stars are shown in
the top and middle panels, respectively, while 35 stars are shown in the bottom
plot.

The abundance patterns in Fig.~\ref{fig:elem} indicate that Fe is a very good
proxy for the metals, since the scatter is very low for the 92 stars plotted.
The $\alpha$ elements show more scatter and, as expected, the low-metallicity
stars have relatively high abundance of these elements.  The CNO abundances are
more uncertain, since they rely on relatively few lines, but they appear to show
the same trend.


It is important to recall that the abundances we have inferred in this work are
based on 1D-LTE model atmospheres of the outer  $\simeq 1$ per~cent of the star.
Furthermore, the content of helium cannot be measured, giving further
uncertainty to the relative abundances. { These surface abundances
are assumed to fully represent those of the stellar interior.} It can be debated
whether the individual abundance patterns of the 13 elements in
Table~\ref{tab:elem} should be used in the asteroseismic modelling.  The scatter
in the individual abundances shown in Fig.~\ref{fig:elem} is relatively small.
Therefore, we recommend a simple approach, based on scaling the solar abundance
with metallicity, except for metal-poor stars.  For ensemble asteroseismology
\citep{chaplin11} we therefore recommend simply to use Fe\,\ione\ (or \fetwo\
for stars hotter than 6000~K) from Table~\ref{tab:elem} to scale all elements. 
However, for stars with metallicity in the range $-0.7 < {\rm [Fe/H]} < -0.2$, 
the $\alpha$ elements should be  increased by 0.15 dex, and by 0.2 dex for CNO. 
{ Population~II stars  (${\rm [Fe/H]} \la -1.0$) are extremely rare in the
\kepler\ field (we only observed one star KIC 8760414) and for these we
recommend using an individually measured  abundance pattern.}


\section{Conclusions}

We have used high-precision stellar spectra to analyse the atmospheric
properties of 93 solar-type stars that have been observed by {\it Kepler}. We
have used the spectra to determine the effective temperature, surface gravity
and heavy-element abundance of each star. These results will facilitate the
asteroseismic  investigation by providing accurate fundamental parameters of
some of the brightest  F5 to K1 type stars of the mission. 

We find that the quantity [Fe/H] is sufficient to describe the metallicity of
most stars in our sample. The [Fe/H] distribution can be well-represented by a
Gaussian with a mean of $-0.06$~dex and FWHM of 0.36~dex.


We confirmed that \logg\ can be determined in a classical spectroscopic analysis
while forcing neutral and ionized lines to give the same abundance. We find a
systematic offset of the spectroscopic and asteroseismic \logg\ of 0.08~dex.
Taking this into account, a spectroscopic \logg\ can be determined with an
accuracy of about 0.10~dex for solar-type stars. 

We note that the {\em PLATO} satellite mission \citep{catala09}, a
candidate M-class mission to the ESA Cosmic Vision program, will
provide asteroseismic parameters for a very large number of
planet-hosting stars. For such a mission, one needs to carefully plan
the huge effort needed to obtain ground-based spectroscopy to
characterize the targets. Since many of the targets will be fainter
than the \kepler\ targets in this work, it will be necessary to work
with low-S/N and/or low-resolution spectra.  It will probably be the
case that our approach to fix \logg\ from the asteroseismic data will
be essential for the spectroscopic analyses.  This could also be true
for the characterization of the many remaining \kepler\ targets with
missing spectroscopic parameters \teff, [Fe/H], and \vsini.

\section*{Acknowledgments}

We are thankful for the efficient service observing teams at the CFHT and Pic
du Midi observatories. Funding for this Discovery
mission is provided by NASA's Science Mission Directorate.
JM-\.Z acknowledges the Polish Ministry grant no N\,N203\,405\,139. 
KU acknowledges financial support by  the Spanish National
Plan of R\&D for 2010, project AYA2010-17803. 
SH acknowledges financial support from the Netherlands Organisation 
for Scientific Research (NWO). 
TSM and HB are supported in part by White Dwarf Research Corporation 
through the Pale Blue Dot project. 
NCAR is sponsored by the U.S. National Science Foundation.
WJC, GAV, YE, and IWR all acknowledge the financial support of
the UK Science and Technology Facilities Council (STFC).
SB acknowledges NSF grant ATM-1105930.

\bibliography{bruntt-v13}
\bsp
\label{lastpage}

\end{document}